\documentclass[conference]{IEEEtran}
\IEEEoverridecommandlockouts

\usepackage{graphicx}
\usepackage{amssymb}
\usepackage{amsmath}
\usepackage{mathptmx}
\usepackage{array}
\usepackage{cite}
\usepackage[hidelinks]{hyperref}
\usepackage{amsthm}

\theoremstyle{plain}
\newtheorem{prop}{Proposition}
\newtheorem{cor}[prop]{Corollary}

\usepackage[shortlabels, inline]{enumitem}

\newcommand{\Trop}{\mathbb{T}}
\newcommand{\eps}{\varepsilon}

\newcommand{\D}{D(S)}
\newcommand{\zeroD}{0_D}
\newcommand{\oneD}{1_D}
\newcommand{\wt}{\operatorname{wt}}
\newcommand{\Id}{\widehat{I}}
\newcommand{\Ahat}{\widehat{A}}
\newcommand{\arrayprod}[2]{\mathbin{{#1}.{#2}}}

\title{{Dual-Number Associative Arrays for One-Pass Provenance and Sensitivity}\\
\thanks{Research was sponsored by the Department of the Air Force Artificial Intelligence Accelerator and was accomplished under Cooperative Agreement Number FA8750-19-2-1000. The views and conclusions contained in this document are those of the authors and should not be interpreted as representing the official policies, either expressed or implied, of the Department of the Air Force or the U.S. Government. The U.S. Government is authorized to reproduce and distribute reprints for Government purposes notwithstanding any copyright notation herein.}
}

\author{\IEEEauthorblockN{
Timothy Habashy, Hayden Jananthan, Jeremy Kepner
\\
\IEEEauthorblockA{
MIT
}}}

\date{}

\begin{document}

\maketitle

\begin{abstract}
Associative arrays and GraphBLAS express sparse computations as matrix operations over a chosen semiring, but the result normally contains only the ordinary value. We define a dual-number extension $D_k(S)=S\ltimes S^k$ that carries this value together with $k$ independent auxiliary channels in the same computation. We prove that the construction is a split square-zero extension of any commutative semiring. Over min-plus, the auxiliary value is the weight of the shortest walk constrained to use the corresponding edge, excluding the weight of that edge. This gives an exact shortest-path membership test for that edge, although ties return the union of the edges on all optimal paths. We implement the min-plus construction using sparse \texttt{python-graphblas} operations, with a fixpoint test that checks the value and auxiliary components together. We validate the implementation across 18,692 checks, including 2,825 tied target pairs. On random graphs with 100,000 vertices and 32 seed edges, the dual method is $3.39\times$ faster than the sequential algorithm but $1.50\times$ slower than a min-plus-specific batched method. This difference is largely explained by the dual method requiring $1.55\times$ as many sweeps. Thus, the main contribution is a general algebraic approach with competitive sparse performance, rather than information that specialized min-plus algorithms cannot already compute.
\end{abstract}

\section{Introduction}

Associative arrays represent spreadsheets, databases, matrices, and graphs as sparse key-labeled arrays whose operations are determined by a chosen value algebra \cite{kepner2015associative}. GraphBLAS applies the same idea to graph algorithms: the sparse matrix structure stays fixed, while addition and multiplication are replaced by semiring operations suited to the problem \cite{kepner2016graphblas,davis2019graphblas}. This provides a compact and reusable language for sparse computation. However, the result usually does not record which inputs contributed to it or how auxiliary information moved through the computation.

Semiring provenance approaches this problem by annotating inputs and propagating those annotations through a computation \cite{green2007provenance,green2017semiring}. Extensions cover aggregate queries \cite{amsterdamer2011aggregate}, linear algebra operators \cite{yan2016linear}, and graph queries \cite{ramusat2021provenance}. Matrix query languages have also been connected to positive relational algebra over semiring-annotated relations \cite{brijder2022matrices}. In ordinary arithmetic, dual numbers carry a value and its first-order information through one forward computation \cite{hoffmann2016ad,vandenberg2024ad}. Expectation semirings similarly carry auxiliary statistics alongside the ordinary value \cite{li2009expectation} and have been studied algebraically as semiring trivial extensions \cite{nasehpour2020expectation}. Newtonian program analysis provides a related differential calculus for semiring polynomials in fixed-point systems \cite{esparza2010newtonian}. This leads to the question studied here: can the dual-number product rule be extended to any commutative semiring, and when does the auxiliary component have a useful interpretation?

This paper makes four contributions: (1) define $D_k(S)=S\ltimes S^k$ as a flat split square-zero extension; (2) interpret its auxiliary components over min-plus; (3) implement a sparse GraphBLAS recurrence with joint convergence and prove termination without reachable negative-weight cycles; and (4) compare sequential and batched min-plus baselines. The batched method recovers the same information at full convergence, but it depends on a min-plus-specific factorization. The dual construction instead uses the same rule for each semiring and works for hop-bounded (restricted to paths containing $\leq N$ edges, where $N$ is fixed) computations.

\section{Preliminaries}

Let $(S,\oplus,\otimes,0_S,1_S)$ be a commutative semiring. Thus, $\oplus$ is associative and commutative with identity $0_S$, while $\otimes$ is associative and commutative with identity $1_S$. Multiplication distributes over addition, and, for all $a \in S$,
\(
0_S\otimes a=a\otimes0_S=0_S
.
\)

For matrices $A$ and $B$ over $S$, we write matrix multiplication as $A \arrayprod{\oplus}{\otimes} B$, with 
\( \displaystyle
(A \arrayprod{\oplus}{\otimes} B)_{ij}=\bigoplus_{\ell} A_{i\ell}\otimes B_{\ell j},
\)
and use matrix powers $A^m$ to denote repeated matrix multiplication under the ambient semiring operations. Using different choices of semirings changes what the same sparse matrix multiplication computes. For example, the plus-times semiring counts walks, while the tropical semiring
\(
\Trop=(\mathbb{R}\cup\{+\infty\},\min,+,+\infty,0)
\)
computes shortest distances \cite{mohri2002semiring}.

Throughout the graph sections in this paper, $n$ denotes the number of vertices of the graph and $N$ denotes a hop budget. A walk from $s$ to $j$ is a sequence $W=(x_0,\ldots,x_h)$ with $x_0=s$, $x_h=j$, and an edge between adjacent nodes in the walk $(x_{\ell-1},x_\ell)$ for every $1\leq\ell\leq h$. Its length (hop count) is $|W|=h$, and its weight is
\(
\wt(W):=\bigotimes_{\ell=1}^{h}w_{x_{\ell-1}x_\ell}=\sum_{\ell=1}^{h}w_{x_{\ell-1}x_\ell}.
\)

\section{The Square-Zero Extension}

\subsection{The rank-one construction}

Let $S$ be a commutative semiring. Define
\(
D=D(S)=\{\,a+b\eps:a,b\in S\,\},
\)
where $\varepsilon$ is a formal symbol that satisfies $\varepsilon^2=0$. Note that $a+b\eps$ can be written in pair-notation as $(a,b)$. The operations are thus
\begin{align*}
(a,b)\oplus(c,d)&=(a\oplus c,b\oplus d),\\
(a,b)\otimes(c,d)
  &=(a\otimes c,(a\otimes d)\oplus(b\otimes c)),
\end{align*}
with $\zeroD=(0_S,0_S)$ and $\oneD=(1_S,0_S)$. This is the semiring form of the Nagata idealization $S\ltimes S$, also called an expectation-semiring or trivial-extension construction \cite{nagata1962local,vandenberg2024ad,nasehpour2020expectation}.

\begin{prop}
\label{proposition 1}
If $S$ is a commutative semiring, then $\D$ is a commutative semiring, and the projection $\phi:\D\to S$, $\phi(a,b)=a$, realizes it as a split square-zero extension:
\[
0\longrightarrow \eps S\longrightarrow \D
\xrightarrow{\ \phi\ }S\longrightarrow0,
\qquad(\eps S)^2=\{\zeroD\}.
\]
\end{prop}

\begin{proof}
Showing that $D(S)$ is a commutative semiring when $S$ is a commutative semiring is a sequence of routine calculations that we omit for brevity.

The semiring laws follow directly from those of $S$. The projection $\phi$ is a surjective homomorphism with section $\iota(a)=(a,0_S)$. Moreover, $I=\{(0_S,b):b\in S\}$ is an ideal and $(0_S,b)\otimes(0_S,d)=\zeroD$, so $I^2=\{\zeroD\}$.
Thus the extension is split and square-zero.
\end{proof}

Note that the product rule also holds at matrix level. If
$A=A_0+A_1\eps$ and $B=B_0+B_1\eps$, then
\(
AB=A_0B_0+
\bigl(A_0B_1\oplus A_1B_0\bigr)\eps.
\)
Thus, one lifted matrix multiplication computes both the ordinary and auxiliary components.

\subsection{Multiple independent markers}

For $k$ markers, construct the $S$-semimodule $S^k$ and define
\[
D_k=D_k(S)=S\ltimes S^k
=\left\{a+\sum_{r=1}^{k}b_r\eps_r\right\},
\]
Where $\eps_1,\dots\varepsilon_k$ are formal symbols that satisfy $\eps_r\eps_\ell=0\quad\forall r,\ell\in \{1,\dots,k\}$. In pair notation with $\mathbf{b}$ and $\mathbf{d}$ as vectors,
\begin{align*}
(a,\mathbf b)\oplus(c,\mathbf d)
&=\bigl(a\oplus c,\mathbf b\oplus\mathbf d\bigr),\\
(a,\mathbf b)\otimes(c,\mathbf d)
&=\bigl(a\otimes c,\,
a\otimes\mathbf d\oplus c\otimes\mathbf b\bigr),
\end{align*}
where the operations and scalar action on $S^k$ are componentwise. The value component and every auxiliary coordinate reduce exactly to the scalar calculations in Proposition~\ref{proposition 1}, so the semiring laws hold coordinate-wise. For $I_k=\{(0_S,\mathbf b):\mathbf b\in S^k\}$,
\(
(0_S,\mathbf b)\otimes(0_S,\mathbf d)=(0_S,\mathbf0),
\)
so $I_k^2=\{0_{D_k}\}$. Projection to the value component is split by
$a\mapsto(a,\mathbf0)$. Also note that the extension must be defined in this flat form; iterating $D(D(\cdots))$ would retain the cross terms $\eps_r\eps_\ell$.

\section{Semiring Instantiations}

\subsection{Min-plus detour distance}

In $D_k(\Trop)$, the operations are componentwise in the auxiliary vector. Thus, focusing on a fixed channel $r$ ($\varepsilon_r$), they reduce to $D_1(\mathbb{T})=D(\mathbb{T})$, where
\(
(a,b)\oplus(c,d) =(\min(a,c),\min(b,d)),
\) and \(
(a,b)\otimes(c,d) =(a+c,\min(a+d,c+b)),
\)
with $\zeroD=(+\infty,+\infty)$ and $\oneD=(0,+\infty)$. For seed edge $(p_r,q_r)$ of weight $w_r$, set its $r$th auxiliary coefficient ($\varepsilon_r$) to $0=1_{\Trop}$; every unseeded coefficient is $+\infty=0_{\Trop}$.

Let $\Ahat$ be this dual-number lift of the weighted adjacency matrix. The dual identity $\Id$ has $(0,+\infty)$ on its diagonal. Then, define
\begin{equation}
\label{equation 2}
B_N:=(\Id\oplus\Ahat)^{ N}
=\bigoplus_{h=0}^{N}\Ahat^{ h},
\end{equation}
which follows from the idempotency of $\oplus$. The right hand equality of \eqref{equation 2} can be proven by induction. We omit for brevity.


Let $\mathcal W_N(s,j)$ be the set of $s\to j$ walks with at most $N$ edges, and let
\(
\mathcal W_{r,N}(s,j)
=\{W\in\mathcal W_N(s,j):W\text{ traverses }(p_r,q_r)\}.
\)
Define
\[
W_{r,N}(s,j)=
\min_{W\in\mathcal W_{r,N}(s,j)}\wt(W),
\]
where every minimum over an empty set is defined as $+\infty$. Write entries of $B_N$ as
\[
(B_N)_{sj}=
\bigl(v_N(s,j),(e_{1,N}(s,j),\ldots,e_{k,N}(s,j))\bigr).
\]
Where $v_N(s,j)$ is the ordinary value and $e_{i,N}(s,j)$ is the coefficient of the $\varepsilon_i$ component of the $s,j$ dual number entry.

\begin{prop}[$\eps$-formula]
\label{proposition 2}
For every $s,j,r$, and $N$,
\[
v_N(s,j)=\min_{W\in\mathcal W_N(s,j)}\wt(W) ~~ \text{and} ~~
e_{r,N}(s,j)=W_{r,N}(s,j)-w_r,
\]
with the convention $+\infty-w_r=+\infty$.
\end{prop}

\begin{proof}
We first prove by induction on $h$ that
\begin{equation}
\label{equation 3}
(\Ahat^{ h})_{sj}
=
\bigoplus_{\substack{W: s\to j\\ |W|=h}}
\bigotimes_{\ell=1}^{h}
\Ahat_{x_{\ell-1}x_\ell}.
\end{equation}
At $h=0$, $\Ahat^{0}=\Id$, which represents the unique $0$-hop empty walk from a vertex to itself and no $0$-hop walk between distinct vertices. Assume \eqref{equation 3} holds for $h$. Matrix multiplication gives
\[
(\Ahat^{(h+1)})_{sj}
=\bigoplus_{u=1}^{n}
(\Ahat^{ h})_{su}\otimes\Ahat_{uj}.
\]
Substituting the induction hypothesis and distributing $\otimes$ over $\oplus$ gives one term for every $h$-hop walk from $s$ to $u$ followed by the edge $(u,j)$. Every $(h+1)$-hop walk has one unique penultimate vertex $u$, so these terms account exactly all $(h+1)$-hop walks from $s$ to $j$. This proves \eqref{equation 3}.

Now fix a nonempty walk $W=(x_0,\ldots,x_h)$ and write the value and $r$th auxiliary components of its $\ell$th edge as
$(\alpha_\ell,\beta_{r,\ell})$. We prove by induction on $h\geq1$ that its dual product has value and $r$th auxiliary component
\begin{equation}
\label{equation 4}
V_h=\sum_{\ell=1}^{h}\alpha_\ell \quad \text{and} \quad 
c_r^{(h)}
=\min_{1\leq\ell\leq h}
\biggl(\beta_{r,\ell}
+\sum_{\substack{1\leq m\leq h \\ m\ne\ell}}\alpha_m\biggr),
\end{equation}
respectively. 
For $h=1$, the product is $(\alpha_1,\beta_{r,1})$; the sum over
$m\ne1$ is empty and equals zero, so \eqref{equation 4} holds. Assume it holds for $h$. Multiplying by the next edge's weight gives value and auxiliary component
\begin{equation} 
\label{equation 5}
V_h+\alpha_{h+1}=\sum_{\ell=1}^{h+1}\alpha_\ell \quad \text{and} \quad
\min\left(V_h+\beta_{r,h+1},
\alpha_{h+1}+c_r^{(h)}\right),
\end{equation}
respectively. 
The first expression in the auxiliary component in \eqref{equation 5} is
\(
\beta_{r,h+1}+\sum_{m=1}^{h}\alpha_m,
\)
which is the $\ell=h+1$ term of \eqref{equation 4}. By the induction hypothesis, the second expression is
\[
\min_{1\leq\ell\leq h}
\biggl(\beta_{r,\ell}
+\sum_{\substack{1\leq m\leq h+1 \\ m\ne\ell}}\alpha_m\biggr),
\]
which contains exactly the terms with $1\leq\ell\leq h$. Their minimum is therefore \eqref{equation 4} with $h+1$ in place of $h$, completing the induction.

The value component is $V_h=\wt(W)$. For channel $r$,
$\beta_{r,\ell}=0$ exactly when the $\ell$th edge is $(p_r,q_r)$ and is $+\infty$ otherwise. If the seeded edge does not occur, every term in \eqref{equation 4} is $+\infty$. If it occurs, then for each seeded edge occurrence,
$\alpha_\ell=w_r$, so its finite term is
\(
0+\sum_{m\ne\ell}\alpha_m=\wt(W)-w_r.
\)
Thus the walk contributes $\wt(W)-w_r$ exactly when it traverses the seeded edge. The empty walk contributes auxiliary value $+\infty$ because the dual multiplicative identity is $(0,+\infty)$.

Finally, \eqref{equation 2} and \eqref{equation 3} show that $(B_N)_{sj}$ is the componentwise $\mathbb{T}$-sum over all walks in $\mathcal W_N(s,j)$. Taking the minimum of their value components gives the formula for $v_N$. Taking the minimum of their $r$th auxiliary components gives
\[
e_{r,N}(s,j)
=\min_{W\in\mathcal W_{r,N}(s,j)}
(\wt(W)-w_r)
=W_{r,N}(s,j)-w_r.
\]
The empty constrained set ($\mathcal W_{r,N}(s,j)=\varnothing$) satisfies the same equality under the stated convention ($+\infty-w_r=+\infty$).
\end{proof}

\begin{cor}
\label{corollary 1}
If $v_N(s,j)<+\infty$, then $(p_r,q_r)$ lies on some minimum-weight $s\to j$ walk of  at most $N$-hops $\iff e_{r,N}(s,j)+w_r=v_N(s,j)$.

Assume no negative-weight cycle is reachable from the source. Under this assumption, this becomes ordinary shortest-path membership.
\end{cor}

\begin{proof}
By Proposition~\ref{proposition 2},
$e_{r,N}(s,j)+w_r=W_{r,N}(s,j)$. Since $\mathcal W_{r,N}(s,j)\subseteq\mathcal W_N(s,j)$,  $W_{r,N}(s,j)$ equals $v_N(s,j)$ exactly when some minimum-weight walk traverses the seeded edge. Under the cycle assumption, this is ordinary shortest-path membership once $N\ge 2n-1$.
\end{proof}

Essentially, $e_{r,N}(s,j)$ is a detour distance through edge $r$ with the weight of $r$ removed. If multiple shortest paths have the same weight, the test identifies an edge whenever it lies on at least one of those shortest paths. Thus, it recovers the union of the edges appearing across all tied shortest paths, but idempotency prevents it from distinguishing which path contributed an edge or how many shortest paths contain it. The hop budget is also part of the output semantics. Let $d(x,y)$ denote unrestricted (no hop bound) shortest-path distance, with
$d(x,y)=+\infty$ when no $x\to y$ walk exists. Then
\(
W_{r,N}(s,j)
\ge d(s,p_r)+w_r+d(q_r,j),
\)
because splitting any forced walk at one seeded edge occurrence gives an
$s\to p_r$ 'prefix' walk, the seeded edge, and a $q_r\to j$ 'suffix' walk. Equality requires shortest choices for both outer portions and enough hops to concatenate them. Consequently, there can be a window $n-1\le N<2n-1$ in which the value channel has converged but an auxiliary channel has not.

\subsection{Other semirings}

The construction is not limited to min-plus. Over plus-times, each auxiliary channel counts traversals of its seeded edge across all fixed-hop walks. Other possible instances include max-plus for critical-path slack, max-min for bottleneck dependence, Boolean channels for dependency masks, and sum-product for expectation statistics. We do not evaluate these additional cases here. As the min-plus example shows, the auxiliary interpretation and any faster specialized factorization must be derived separately for each semiring.

\section{Implementation and Convergence}

We implement the recurrence in \texttt{python-graphblas}, a Python interface to SuiteSparse:GraphBLAS \cite{pythonGraphblas2025,davis2019graphblas}. The value channel is a sparse vector $v^{(t)}\in\Trop^n$. The $k$ auxiliary channels are packed as a sparse matrix $E^{(t)}\in\Trop^{k\times n}$. We merge a value-only zero diagonal into $A$ and leave the auxiliary diagonal structurally absent.

For seed edge $(p_r,q_r)$, define $J(v)_{r, j}$ to be $v_{p_r}$ if $j = q_r$ and $+\infty$ otherwise.
Before the value vector is frozen, one sweep is
\[
v^{(t+1)}=v^{(t)}\arrayprod{\oplus}{\otimes} A,\qquad
E^{(t+1)}=(E^{(t)}\arrayprod{\oplus}{\otimes} A)\oplus J(v^{(t)}).
\tag{6}
\]
Initialize $v^{(0)}_s=0$, $v^{(0)}_j=+\infty$ for $j\ne s$, and
$E^{(0)}_{r,j}=+\infty$. Equation (6) is exactly multiplication of the
lifted source row by $\Id\oplus\Ahat$; hence, after $t$ sweeps, the state
equals that row multiplied by $B_t$ and has the at-most-$t$-hop semantics of Proposition~\ref{proposition 2}.

The first product is a sparse vector-matrix multiplication, and the
second is a sparse matrix-matrix multiplication. The implementation
uses double buffers and an $O(k)$ injection loop, with no Python loop
over vertices. Once $v^{(t+1)}=v^{(t)}$, we freeze $v$ and omit later
injections. Indeed, $J(v^{(t+1)})=J(v^{(t)})$ and
$E^{(t+1)}\leq J(v^{(t)})$. Since $A_{jj}=0$,
$E^{(t+1)}\arrayprod{\oplus}{\otimes} A\leq E^{(t+1)}$; induction therefore shows that
later injections cannot improve $E$.

\begin{prop}
\label{proposition 3}
If one complete sweep leaves both $v$ and $E$ unchanged, every later sweep also leaves them unchanged.
\end{prop}

\begin{proof}
After $v$ is frozen, the recurrence is $E^{(t+1)}=E^{(t)}\arrayprod{\oplus}{\otimes} A$. Therefore, if one propagation leaves $E$ unchanged, every subsequent propagation has the same input and
output.
\end{proof}

\begin{prop}
\label{proposition 4}
Assume no negative-weight cycle is reachable from the source. Every value and auxiliary entry reaches its final value by sweep $2n-1$, and the algorithm terminates by sweep $2n$.
\end{prop}

\begin{proof}
Consider a finite $s\to j$ walk that repeats a vertex. The segment between two consecutive occurrences of that vertex is a directed cycle reachable from $s$, so its weight is nonnegative. Removing that segment leaves an $s\to j$ walk of no greater weight and with fewer edges. Repeating this operation produces a simple walk. Thus every $s\to j$ walk has a simple representative of no greater weight. There are only finitely many simple walks, so if $j$ is reachable, the least weight among them exists and is also the unrestricted shortest-path weight. A simple walk has at most $n-1$ edges. If $j$ is unreachable, its value remains $+\infty$. Hence every value entry is final by sweep $n-1$.

Now fix seed edge $(p_r,q_r)$ and target $j$. If no $s\to j$ walk traverses the seeded edge, the corresponding auxiliary entry remains $+\infty$ and is already final. Otherwise, take any forced walk $W$ and choose one occurrence of the seeded edge. Its prefix ends at $p_r$ and its suffix begins at $q_r$, so
\begin{equation} \label{equation 8}
\wt(W)\geq d(s,p_r)+w_r+d(q_r,j),
\end{equation}
where $d$ is unrestricted shortest-path distance. Because a forced walk exists, $q_r$ is reachable from $s$ through the seeded edge. Thus every cycle in either outer portion is reachable from $s$ and has nonnegative weight. The cycle-removal argument above gives a shortest $s\to p_r$ walk and a shortest $q_r\to j$ walk, each with at most $n-1$ edges. Concatenating them with the seeded edge produces a forced walk whose weight equals the right side of \eqref{equation 8}. Hence it is minimum-weight and uses at most
\(
(n-1)+1+(n-1)=2n-1
\)
edges.

By Proposition~\ref{proposition 2}, the auxiliary entry at sweep $N$ is the minimum over forced walks of at most $N$ edges, hence every auxiliary entry is final by sweep $2n-1$. All value entries are also final by then. Sweep $2n$ cannot lower either component because each already equals its unrestricted minimum, so is unchanged and the joint fixpoint test terminates the algorithm.
\end{proof}

\section{Evaluation}

\subsection{Correctness and protocol}

We first compared the min-plus implementation with a NetworkX oracle on graphs with unique and tied shortest paths. The suite checked the ordinary value, every detour channel, and the on-path predicate across 18,692 outputs, including 2,825 tied target pairs. All checks agreed. Before timing, we also required exact agreement among the complete distance and detour arrays from all three methods, including the locations of $+\infty$ entries.

We timed three methods: \begin{enumerate*}[(1)] \item the coupled one-pass dual recurrence; \item $k+1$ sequential single-source solves, one from $s$ and one from each $q_r$; and \item one batched $(k+1)\times n$ GraphBLAS solve. \end{enumerate*} At full convergence, the batched baseline uses
\(
d(s,p_r)+w_r+d(q_r,j)
\)
to recover each forced distance. This gives the batched method access to the same fully converged detour information and provides a demanding min-plus baseline. The factorization is also closely related to conventional shortest-path sensitivity analysis \cite{ramaswamy2005sensitivity}.

We generated random directed graphs with positive integer weights from 1 to 10 and average out-degree 8. The $k$-scaling experiment fixed $n=100{,}000$ and used $k=2,4,6,\ldots,32$. The $n$-scaling experiment fixed $k=8$ and used 10,000 through 200,000 vertices. Each point is the median of three trials, with first-to-third-quartile error bars. We ran the experiments with Python 3.13.7, \texttt{python-graphblas} 2025.2.0, and 14 GraphBLAS threads on an Apple arm64 system.

\begin{figure*}[htbp]
\centering
\begin{minipage}[htbp]{0.45\textwidth}
\centering
\includegraphics[width=\linewidth]{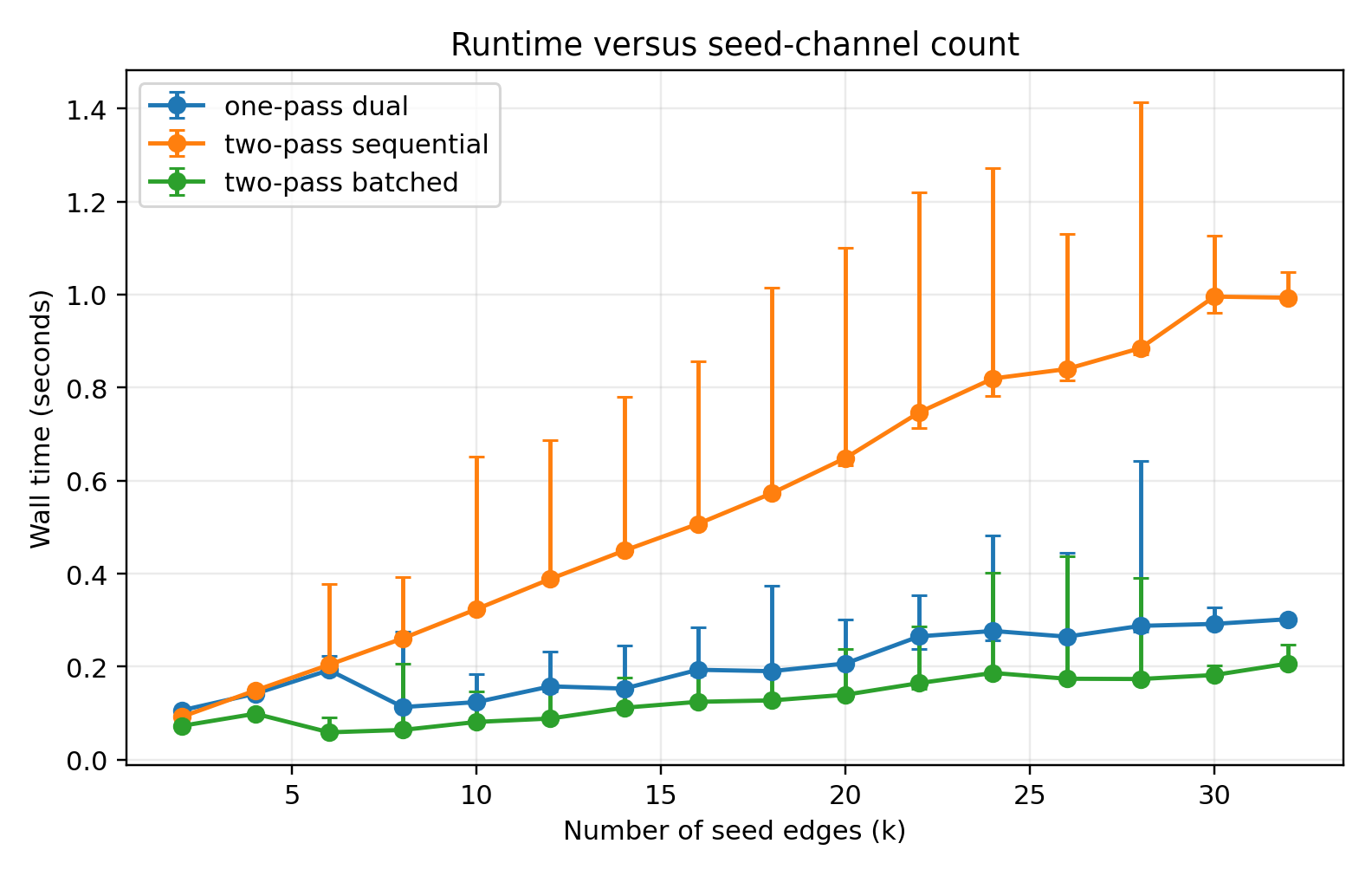}\\[-1mm]
\small (a)
\end{minipage}
\hfill
\begin{minipage}[htbp]{0.45\textwidth}
\centering
\includegraphics[width=\linewidth]{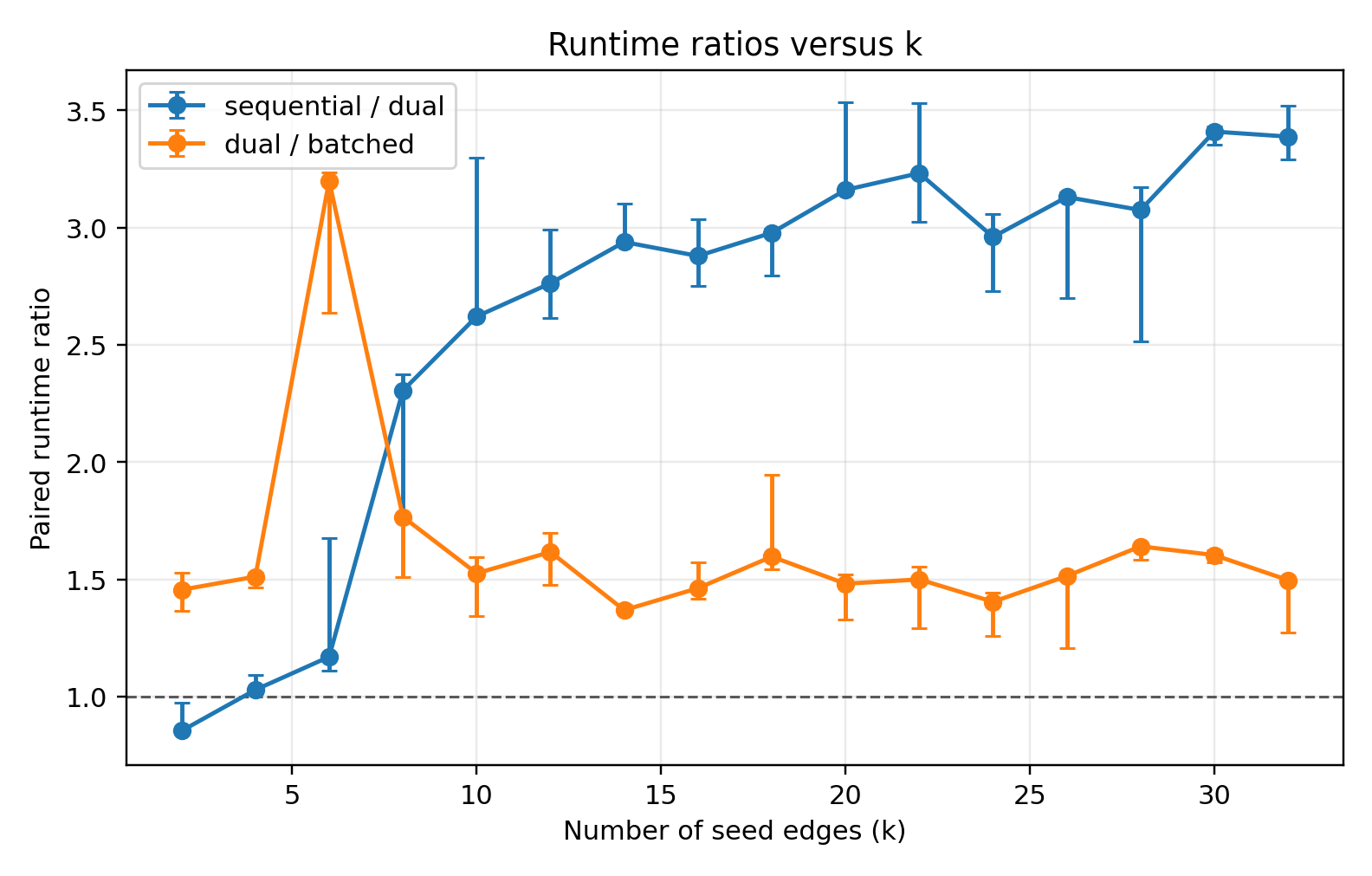}\\[-1mm]
\small (b)
\end{minipage}
\caption{Seed-channel scaling at $n=100{,}000$. (a) Median wall time for the dual, sequential, and batched methods. (b) Paired runtime ratios. A ratio above 1 favors the denominator method. Error bars show the first and third quartiles.}
\label{fig:k-scaling}
\end{figure*}

\begin{figure*}[htbp]
\centering
\begin{minipage}[htbp]{0.45\textwidth}
\centering
\includegraphics[width=\linewidth]{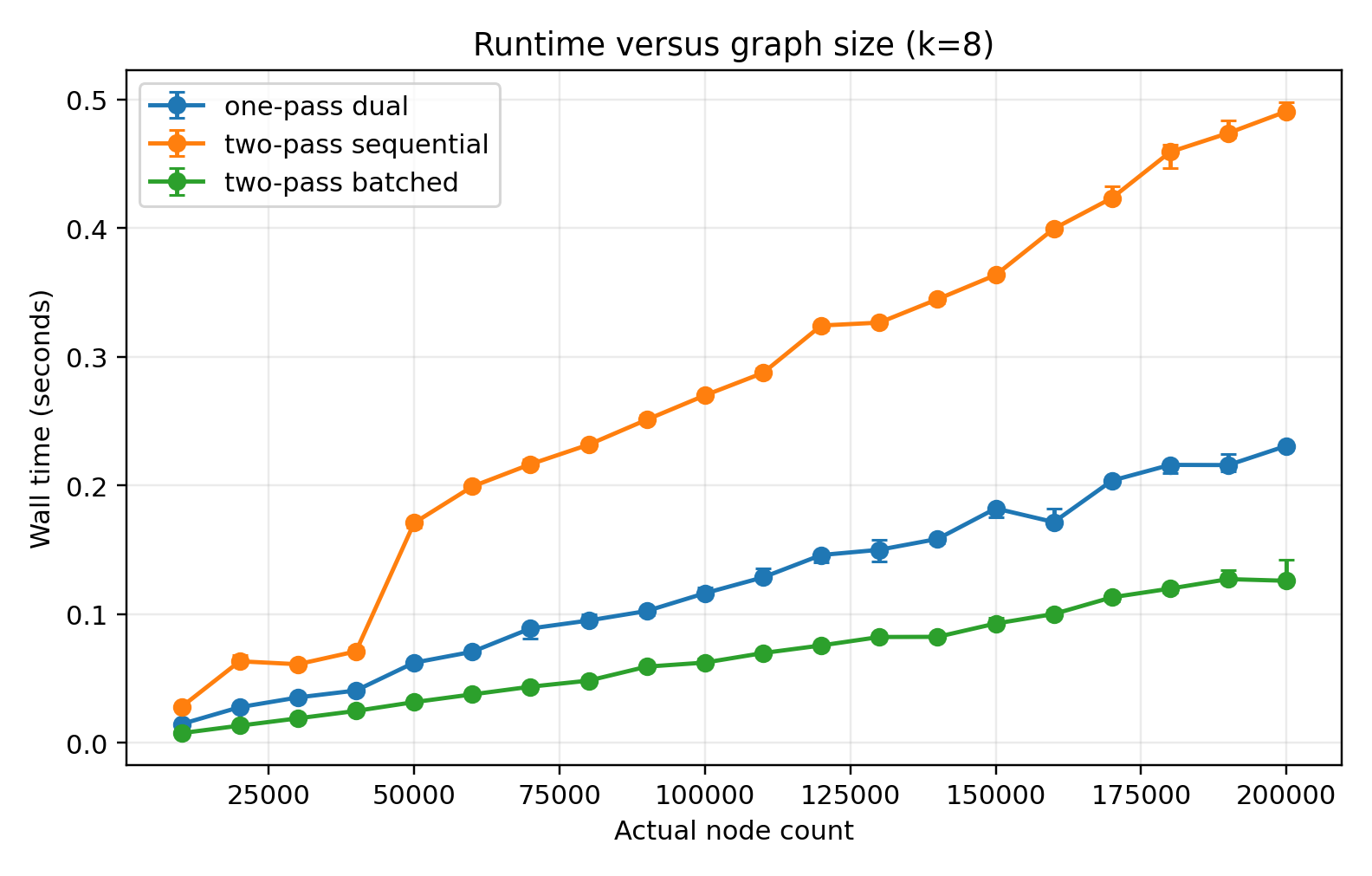}\\[-1mm]
\small (a)
\end{minipage}
\hfill
\begin{minipage}[htbp]{0.45\textwidth}
\centering
\includegraphics[width=\linewidth]{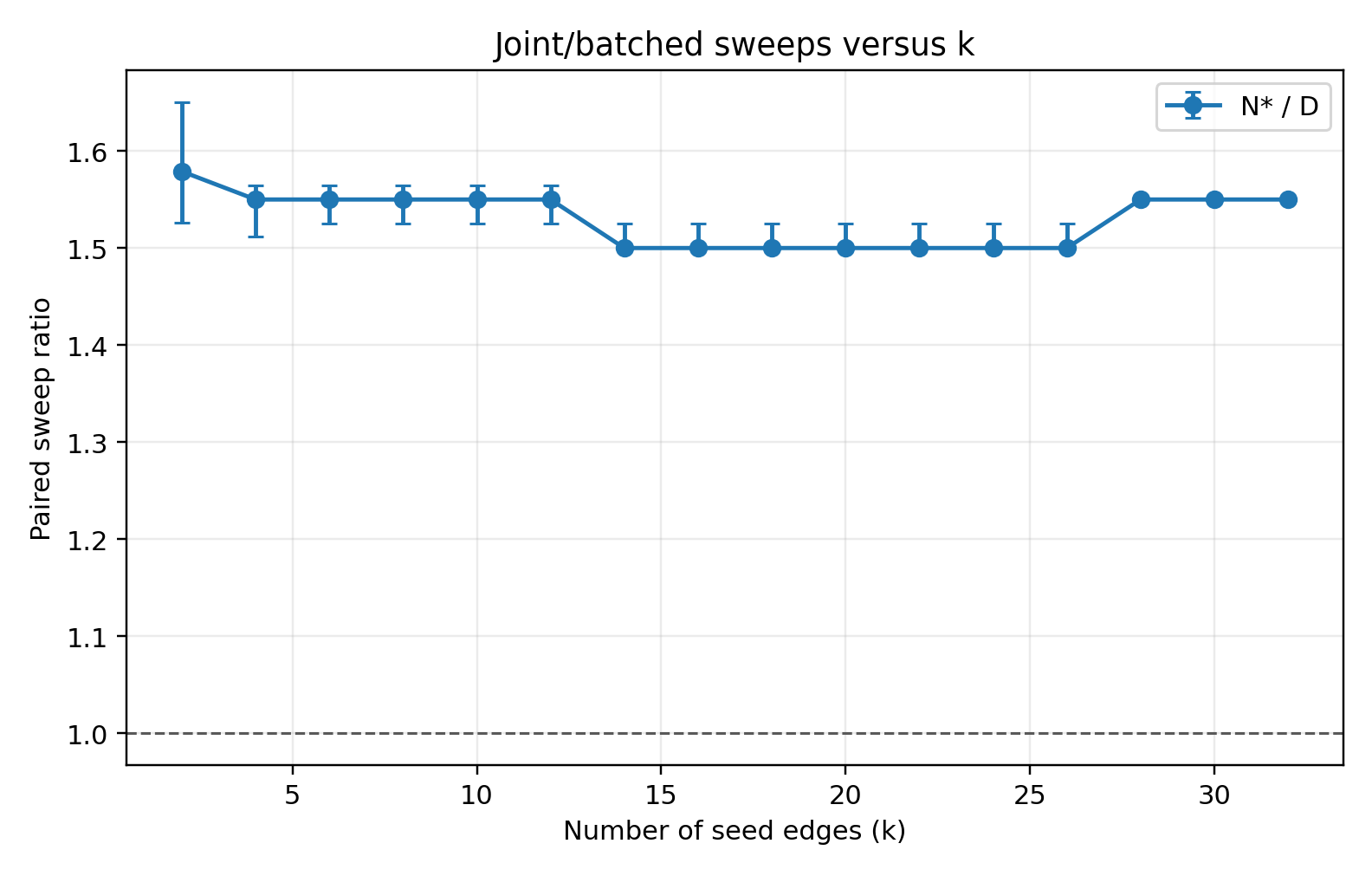}\\[-1mm]
\small (b)
\end{minipage}
\caption{Graph-size and convergence results. (a) Runtime versus actual vertex count at $k=8$. (b) The dual joint-fixpoint sweep count $N^\ast$ divided by the batched distance sweep count $D$.}
\label{fig:n-scaling}
\end{figure*}

\subsection{Results}

As shown in Fig.~\ref{fig:k-scaling}, the sequential runtime increased most quickly with $k$. At $k=32$, its median runtime was 0.993 s, compared with 0.302 s for the dual method and 0.207 s for the batched method. On paired trials, the dual method was $3.39\times$ faster than sequential and $1.50\times$ slower than batched. It was slower than sequential at $k=2$, approximately tied at $k=4$, and grew faster as the number of seed channels grew.

The batched method gains its advantage from the min-plus factorization after ordinary distance convergence. The dual recurrence instead propagates complete forced walks, which can be nearly twice as long as ordinary shortest paths. Across the $k$ experiment, $N^\ast/D$ remained between approximately 1.50 and 1.58. At $k=32$, it was 1.55, close to the 1.50 runtime ratio.

Runtime also grew approximately linearly over the graph sizes in Fig.~\ref{fig:n-scaling}. At $n=200{,}000$ and $k=8$, median runtimes were 0.231 s for dual, 0.491 s for sequential, and 0.126 s for batched. At full convergence, the dual extension neither adds min-plus detour information nor outperforms the specialized batched method. Its contribution is instead a semiring-generic algebraic framework that propagates auxiliary information uniformly across semiring choices while preserving direct $N$-hop semantics, properties that the min-plus-specific full-distance factorization does not provide.

The timing study uses synthetic graphs, three trials per point, and one local system. We therefore do not yet make a hardware-independent scaling claim.

\section{Conclusion}

We defined $D_k(S)=S\ltimes S^k$ as a flat split square-zero extension that carries one ordinary value and $k$ independent auxiliary channels through a sparse semiring computation. In min-plus, it gives a forced-through detour distance and an exact shortest-path membership test, with the stated limits under tied paths and finite hop budgets. Under the cycle assumption of Proposition~\ref{proposition 4}, the packed GraphBLAS implementation reaches a jointly confirmed fixpoint by sweep $2n$. Though the experiments show a clear tradeoff---as $k$ grows, the dual method becomes faster than sequential solves but remains slower than the min-plus-specific batched solver because forced walks require more sweeps---this still exhibits a win for the proposed method since it is not substantially less performant and provides the user a more robust output.

\section*{Acknowledgment}
The authors wish to acknowledge the following individuals for their contributions and support 
L. Anderson, W. Arcand, D. Bestor, W. Bergeron, B. Bond, C. Byun, A. Bonn, D. Burrill, V. Gadepally, J. Gottschalk, T. Hardjono, M. Houle, M. Hubbell, M. Jones, P. Luszczek, P. Michaleas, L. Milechin, J. Mullen, C. Leiserson, C. Milner, S. Mohindra,  A. Pentland, A. Prout, C. Prothmann, A. Reuther, A. Rosa, D. Rus, D. Ross, J. Ross, M. Sherman, S. Van Broekhoven, M. Weems, J. Wilkinson, C. Yee.

\bibliographystyle{IEEEtran}
\bibliography{references}

\end{document}